# Mechanisms for Multi-Unit Auctions

**Shahar Dobzinski**                                    SHAHAR@CS.CORNELL.EDU
*Computer Science Department, Cornell University*
*Ithaca, NY 14853*

**Noam Nisan**                                          NOAM@CS.HUJI.AC.IL
*The School of Computer Science and Engineering*
*The Hebrew University of Jerusalem*
*Jerusalem, Israel*

## Abstract

We present an incentive-compatible polynomial-time approximation scheme for multi-unit auctions with general $k$-minded player valuations. The mechanism fully optimizes over an appropriately chosen sub-range of possible allocations and then uses VCG payments over this sub-range. We show that obtaining a *fully* polynomial-time incentive-compatible approximation scheme, at least using VCG payments, is NP-hard. For the case of valuations given by black boxes, we give a polynomial-time incentive-compatible 2-approximation mechanism and show that no better is possible, at least using VCG payments.

## 1. Introduction

In Algorithmic Mechanism Design our goal is to construct efficient mechanisms that will handle the selfish behavior of the players. In particular, we are interested in designing *truthful* mechanisms, that is, mechanisms in which the dominant strategy of each player is to simply reveal his true valuation.

Typical problems in the field involve allocating goods to players. One key problem is the problem of multi-unit auctions. Here we are given $m$ identical items and $n$ bidders. In our setting we view the number of items $m$ as "large" and desire mechanisms whose computational complexity is polynomial in the *number of bits* needed to represent $m$. Every bidder $i$ has a valuation function $v_i : \{1, ..., m\} \rightarrow \Re$, where $v_i(q)$ denotes his value for obtaining $q$ items. We assume that $v_i$ is weakly monotone increasing (free disposal), and normalized ($v_i(0) = 0$). Our goal is the usual one of maximizing the social welfare $\Sigma_i v_i(q_i)$ where $\Sigma_i q_i \leq m$.

In the general case, each $v_i$ is represented by $m$ real numbers, and in abstract settings may be accessed as a black box. In a concrete setting, we will assume that $v_i$ is represented as a $k$-minded bid, i.e. given by a list: $(q_1, p_1), ..., (q_k, p_k)$, where $v_i(q) = \max_{\{j|q_j \leq q\}} p_j$. This corresponds to a XOR bidding language (Nisan, 2006). Observe that a $k$-minded valuation corresponds to a step function with at most $k$ steps (step $i$ is located at $q_i$ and has height $p_i$). In general, $k$ can be as large as $m$, and the case $k = 1$ is the single-minded case.

This problem has received much attention, starting from Vickrey's seminal paper (1961) that described a truthful mechanism for the case of "downward sloping valuations" in which the items can be optimally allocated greedily. The general case, however, is NP-hard, as the single-minded case is just a re-formulation of the knapsack problem. Luckily, the knapsack problem has a fully-polynomial time approximation scheme (i.e. can be approximated to within a factor of $1 + \epsilon$ in time polynomial in $n, \log m, \epsilon^{-1}$), and it is not hard to see that the algorithm directly extends to the general case of multi-unit auctions.

## 1.1 VCG-Based Mechanisms

The key positive technique of mechanism design is the VCG payment scheme. In this payment scheme each bidder $i$ pays $h_i(v_{-i}) - \Sigma_{j \neq i} v_j(a)$, where $a$ is the algorithmic output, and $h_i$ is some arbitrary function that does not depend on $v_i$. Unfortunately, while VCG works perfectly well from a game-theoretic perspective, it is not so useful in computational settings, since in multi-unit auctions, and in "most" interesting combinatorial optimization problems, calculating the exact optimum is intractable.

One naive idea is to use an approximation algorithm to find an approximate solution $a$, and then let each bidder $i$ pay $h_i(v_{-i}) - \Sigma_{j \neq i} v_j(a)$. Applying this idea to the case of multi-unit auction is tempting in particular in light of the known $(1 + \epsilon)$-approximation algorithm for this problem. Unfortunately, it turns out that in general using an approximation algorithm together with VCG payments does not result in a truthful mechanism. This phenomenon was studied by Nisan and Ronen (2007). It was observed there that the following family of allocation algorithms do yield truthful VCG-based mechanisms:

**Definition**: An allocation algorithm (that produces an output $a \in \mathcal{A}$ for each input $v_1, ..., v_n$, where $\mathcal{A}$ is the set of possible alternatives) is called "maximal-in-range" (henceforth MIR) if it completely optimizes the social welfare over some subrange $\mathcal{R} \subseteq \mathcal{A}$. I.e., for some $\mathcal{R} \subseteq \mathcal{A}$, we have that for all $v_1, ..., v_n$, $a \in \arg\max_{a \in \mathcal{R}} \Sigma_i v_i(a)$.

I.e., MIR algorithms use the following natural and simple strategy to find an approximately optimal solution: they just optimally search within a pre-specified sub-range of feasible solutions – a subrange over which optimal search is algorithmically feasible.

The main result of Nisan and Ronen (2007) states that this is essentially it – no other VCG-based mechanisms are incentive compatible.

**Theorem (Nisan & Ronen, 2007):** The allocation algorithm of any incentive-compatible VCG-based mechanism for combinatorial auctions is equivalent to a maximal-in-range algorithm.

"Equivalent" here means that the social utilities are identical for all inputs, i.e. if $a$ and $b$ are the outputs of the two allocation algorithms for input $v_1, ..., v_n$ then $\Sigma_i v_i(a) = \Sigma_i v_i(b)$. In particular, the outputs must coincide generically – except perhaps in case of ties. Nisan and Ronen (2007) prove this for two specific types of mechanism design problems, but the result is more general (Dobzinski & Nisan, 2007).

Following Nisan and Ronen (2007), Dobzinski and Nisan (2007) view this result as a negative one. Namely, they show that in the setting of combinatorial auctions with submodular bidders, MIR algorithms do not have much power. This might imply that in the above setting truthful deterministic mechanisms do not have much power, since Lavi et al. (2003) give a partial evidence that all truthful mechanisms that provide a good approximation ratio *must* be MIR.

In contrast, this paper views this result as a positive result. We observe that MIR algorithms provide us with a constructive way of obtaining truthful mechanisms[1]. Indeed, this paper suggests that there are some settings where the power of MIR algorithms is not trivial at all. We note that several previous papers already obtained upper bounds using MIR techniques (Holzman, Kfir-Dahav, Monderer, & Tennenholtz, 2004; Dobzinski, Nisan, & Schapira, 2005; Blumrosen & Dobzinski, 2007). Yet, this paper initiates the systematic study of MIR algorithms. In particular, the techniques used are more sophisticated then those obtained in previous work.

## 1.2 Previous Work and Our Results

For the case of multi-unit auctions with single-minded bidders, the paper by Briest, Krysta, and Vöcking (2005) presents a truthful fully polynomial time approximation scheme (FP-TAS), improving upon a previous result of Mu'alem and Nisan (2002). The only result known for $k$-minded bidders is a randomized $\frac{1}{2}$-approximation mechanism that is truthful in expectation (Lavi & Swamy, 2005). The current paper presents a polynomial time approximation scheme (PTAS) for the general case.

**Theorem:** For every fixed $\epsilon > 0$, there exists a truthful $(1 - \epsilon)$-approximation mechanism for multi-unit auctions with $k$-minded bidders whose running time is polynomial in $n$, $\log m$, and $k$. The dependence of the running time on $\epsilon$ is exponential.

However, we prove two ways in which the mechanism can not be improved upon. First, we show that the dependence on $\epsilon$ cannot be made polynomial without destroying the truthfulness, as long as MIR techniques are used. In contrast, from a pure algorithmic point of view it is possible to obtain a fully polynomial time approximation scheme[2]. Furthermore, there exists a truthful FPTAS if all bidders are known to be single minded (Briest et al., 2005).

**Theorem:** No fully polynomial time truthful approximation mechanism that uses VCG payments exists, unless P=NP.

Then we show that the dependence on $k$ is necessary, and that no approximation scheme is possible in a general black-box model. This is shown in a general communication model, and even for two bidders.

**Theorem:** Every approximation mechanism among two bidders with general valuations that uses VCG payments requires exponentially many queries to obtain an approximation factor that is better than $\frac{1}{2}$.

We do present a truthful approximation mechanism in the general black box model that does obtain a factor of $\frac{1}{2}$. This improves upon the randomized one of Lavi and Swamy (2005) that is only truthful in expectation.

---

1. Note that the payments of an efficient MIR mechanism can be computed efficiently: take the output allocation and pay each bidder the sum of the values of the other bidders in the output allocation.

2. Of course, ignoring computational issues, the standard VCG mechanism is MIR and provides an approximation ratio of 1.

**Theorem:** There exists a truthful $\frac{1}{2}$-approximation mechanism for multi-unit auctions among general valuations whose running time is polynomial in $n$ and $\log m$. The access to bidders' valuations is through value queries: "what is $v_i(q)$?"[3]. The mechanism uses VCG payments.

In Section 5 we present a fairly general construction that takes *any* $\alpha$-approximation MIR algorithm $A$ for $t$ bidders, for some $\alpha \leq 1$, and converts it into an $(\alpha - \frac{1}{t+1})$-approximation algorithm for $n$ bidders, in time polynomial in $n$, $\log m$, and the running time of $A$. The construction works as long as the model allows us to answer value queries. We present four applications of the construction: the first two applications provide another proof of the upper bounds discussed above (the PTAS for $k$-minded bidders, and the $\frac{1}{2}$-approximation for the black-box model). Then we present two new applications: a PTAS for stronger bidding languages, such as the one used by Kothari et al. (2005), and a $(\frac{3}{4} + \epsilon)$-approximation mechanism for multi-unit auctions with subadditive valuations. Prior to our paper, nothing was known about the latter setting.

The construction provides us with an interesting example of a *truthful reduction* among problems: any MIR approximation algorithm for a fixed number of bidders can be automatically translated into a truthful approximation algorithm for any number of bidders, while losing only a small factor in the approximation ratio.

## 1.3 Discussion and Open Questions

The main open problem remains to determine the best approximation ratio that can be obtained in a truthful way. The only general method known for constructing such mechanisms is VCG, and our lower bounds state that VCG cannot take us any further. Furthermore, Lavi et al. (2003) show that for 2 bidders in the black-box model, *if all items are allocated*, then MIR algorithms are the only truthful way to obtain a reasonable approximation ratio. Combined with our lower bounds, we get that no better-than-$\frac{1}{2}$ truthful approximation is possible in polynomial time, for 2 bidders and if all items are allocated. An intriguing open question is to understand whether the condition that all items are allocated is indeed necessary.

Another issue that the paper highlights is the inherent difference between obtaining approximation algorithms in single-parameter environments and multi-parameter environments. In a single parameter environment the private information of each bidder consists of only one number, while in multi-parameter environments the private information consists of more than one number. Recall that in the single parameter variant of multi-unit auctions, where we assume that all bidders are single-minded, there exists a truthful FPTAS (Briest et al., 2005). However, the variants we discuss in this paper are multi-parameter, and indeed our lower bounds suggest that obtaining FPTAS is impossible (if one can prove that all mechanisms that give a good approximation ratio must be MIR).

## Paper Organization

In Section 3 we present the PTAS for $k$-minded bidders, and the $\frac{1}{2}$-approximation in the black-box model. Section 4 considers lower bounds for MIR algorithms in both models.

---

3. This is yet another improvement upon the mechanism of Lavi and swamy (2005) which requires the stronger *demand* queries.

In Section 5 we describe a general construction, and its algorithmic applications: truthful mechanisms for other models and more powerful bidding languages.

## 2. Preliminaries

In this section we provide the basic definitions and notations used in this paper.

### 2.1 The Setting

In a multi-unit auction there is a set of $m$ identical items, and a set $N = \{1, 2, \ldots, n\}$ of bidders. Each bidder $i$ has a valuation function $v_i : [m] \to \mathbb{R}^+$, which is normalized ($v_i(0) = 0$) and non-decreasing. Denote by $V$ the set of possible valuations. An allocation of the items $\vec{s} = (s_1, \ldots, s_n)$ to $N$ is a vector of non-negative integers with $\Sigma_i s_i \leq m$. Denote the set of allocations by $S$. The goal is to find an allocation that maximizes the welfare: $\Sigma_i v_i(s_i)$.

We consider two settings that differ in how the valuations are given to us. In the concrete setting (a "bidding language" model) we will assume that $v_i$ is represented as a $k$-minded bid, i.e. given by a list: $(q_1, p_1), \ldots, (q_k, p_k)$, where $v_i(q) = \max_{\{j | q_j \leq q\}} p_j$.

Otherwise, the valuations are given to us as black boxes. For algorithms, the black box $v$ will only answer the weak value queries: given $s$, what is the value of $v(s)$. For the impossibility result, we assume that the black box $v$ can answer any query that is based on $v$ (the "communication model"). Our algorithms run in time $poly(n, \log m)$, while our impossibility result gives a lower bound on the number of bits transferred, and holds even if the mechanism is computationally unbounded.

### 2.2 Truthfulness

A deterministic $n$-bidder mechanism for multi-unit auctions is a pair $(f, p)$ where $f : V^n \to S$ and $p = (p_1, \cdots, p_n)$, where $p_i : V^n \to \mathbb{R}$.

**Definition 2.1** *Let $(f, p)$ be a deterministic mechanism. $(f, p)$ is truthful if for all $i$, all $v_i, v_i'$ and all $v_{-i}$ we have that $v_i(f(v_i, v_{-i})_i) - p_i(v_i, v_{-i}) \geq v_i(f(v_i', v_{-i})_i) - p(v_i', v_{-i})$.*

**Definition 2.2** *$f$ is an affine maximizer if there exist a set of allocations $\mathcal{R}$, a constant $\alpha_i \geq 0$ for $i \in N$, and a constant $\beta_{\vec{s}} \in \Re$ for each $\vec{s} \in S$, such that $f(v_1, ..., v_n) \in \arg\max_{\vec{s}=(s_1,...,s_n) \in \mathcal{R}} (\Sigma_i(\alpha_i v_i(s_i)) + \beta_s)$. $f$ is called maximal-in-range (MIR) if $\alpha_i = 1$ for $i \in N$, and $\beta_s = 0$ for each $\vec{s} \in \mathcal{R}$.*

The following proposition is standard:

**Proposition 2.3** *Let $f$ be an affine maximizer (in particular, maximal in range). There are payments $p$ such that $(f, p)$ is a truthful mechanism.*

## 3. The Basic Mechanisms

This section provides the basic mechanisms for multi-unit auctions: a PTAS for $k$-minded bidders, and a $\frac{1}{2}$ approximation for the black-box model.

### 3.1 A Truthful PTAS for $k$-Minded Bidders

We design an MIR algorithm for this problem, which directly yields an incentive-compatible VCG-based mechanism. We define a range $\mathcal{R}$ of allocations, and prove that if all bidders are $k$-minded then the algorithm outputs in polynomial time the best allocation in $\mathcal{R}$. We start with defining the subrange $\mathcal{R}$.

**Definition 3.1** *We say that an allocation $(s_1, ..., s_n)$ is $t$-round if there exists a set $T$ of bidders, $|T| \leq t$, such that the following two conditions hold:*

- *Let $l = \Sigma_{j \in T} s_j$.*

- *For each bidder $i \notin T$, $s_i$ is a multiple of $\max(\left\lfloor \frac{m-l}{(n-t)^2} \right\rfloor, 1)$.*

- *At most $\max(\left\lfloor \frac{m-l}{(n-t)^2} \right\rfloor, 1) \cdot (n-t)^2$ items are assigned to bidders outside $T$: $\Sigma_{i \notin T} s_i \leq \max(\left\lfloor \frac{m-l}{(n-t)^2} \right\rfloor, 1) \cdot (n-t)^2$*

We let $\mathcal{R}$ be the set of all $t$-round allocations for some fixed $t$ (that will depend only on the approximation guarantee). Next we prove that the value of the best allocation in $\mathcal{R}$ is close to the optimum:

**Lemma 3.2** *Let $(a_1, ..., a_n)$ be an optimal $t$-round allocation, and $(o_1, ..., o_n)$ an optimal unrestricted allocation, then $\Sigma_i v_i(a_i) \geq (1 - \frac{1}{t+1})\Sigma_i v_i(o_i)$.*

**Proof:** Let us start with an optimal unrestricted allocation $(o_1, ..., o_n)$, and use it to construct a $t$-round allocation with high value. Assume that all items are allocated in the optimal allocation (without loss of generality due to the monotonicity of the valuations), and that $v_1(o_1) \geq ... \geq v_n(o_n)$. Let $T = \{1, ..., t\}$ be the set of $t$ bidders required in Definition 3.1, and assign each bidder $i \in T$ a bundle of size $o_i$. As in Definition 3.1, let $l = \Sigma_{j \in T} o_j$.

Let $j \notin T$ be the bidder who got the largest number of items $o_j \geq \frac{m-l}{n-t}$. For each $i \notin T$, $i \neq j$, round *up* each $o_i$ to the nearest multiple of $b = \max(\left\lfloor \frac{m-l}{(n-t)^2} \right\rfloor, 1)$ and assign this many items to bidder $i$. Assign bidder $j$ no items. This is a valid $t$-round allocation since if $b \neq 1$ we added at most $(n-t) \cdot \left\lfloor \frac{m-l}{(n-t)^2} \right\rfloor \leq \frac{m-l}{n-t}$ items by rounding up, but deleted at least $\frac{m-l}{n-t}$ items by removing $o_j$. Notice that the second condition also holds. If $b = 1$, observe that even the optimal allocation is $t$-round. As for the value of the solution, observe that each bidder $i \neq j$ gets a bundle no smaller than $o_i$. In addition, $v_j(o_j) \leq \frac{\Sigma_i v_i(o_i)}{t+1}$, which gives the required approximation. $\qquad \square$

Our MIR approximation algorithm will try each subset of at most $t$ bidders to be the set $T$ of bidders. For each possible selection of $T$, the algorithm will consider all possible allocations to bidders in $T$ according to the $k$ bids each bidder submitted. That is, we will consider the allocation that assigns each bidder $i \in T$ exactly $s_i$ items, if and only if $\Sigma_{i \in T} s_i \leq m$, and for each $s_i$ there is a bid $(s_i, p_i)$ in the $k$ bids of bidder $i$ (for some $p_i > 0$).

For each selection of $T$ and allocation to the bidders in $T$ according to their bids, the algorithm splits the remaining $m - l$ items into at most $(n-t)^2$ equi-sized bundles of size $\max(\left\lfloor \frac{m-l}{(n-t)^2} \right\rfloor, 1)$, where $l$ is the total number of items that bidders in $T$ get. The maximal-in-range algorithm will optimally allocate these equi-size bundles among the bidders that are not in $T$. Finally, the algorithm outputs the best allocation among all allocations considered. All that is left is to show the following two lemmas:

**Lemma 3.3** *For every fixed $t$ the above algorithm runs in time polynomial in $n$ and $\log m$.*

**Proof:** There are at most $n^t$ possible selections of sets $T$. For each selection of $T$ there are at most $k^t$ allocations to bidders in $T$ that are considered. Finding the optimal allocation to bidders not in $T$ is by dynamic programming. Let $b$ be the size of the equi-size bundles. Without loss of generality, we assume that $T = \{n - t + 1, ..., n\}$. We calculate the following information for every $1 \leq i \leq n - t$ and $1 \leq q \leq (n - t)^2$: $M(i, q)$ is the maximum value that can be obtained by allocating at most $q$ equi-size bundles among bidders $1...i$. Each entry can be filled in polynomial time using the relations: $M(i, q) = \max_{q' \leq q}(v_i(q'b) + M(i - 1, q - q'))$. In particular notice that if $b = 1$ then the number of equi-size bundles is polynomial in the number of bidders, thus the number of entries in the table is polynomial also in this case. Overall we get that the algorithm runs in time polynomial in $n$ and $\log m$, for every fixed $t$. $\square$

**Lemma 3.4** *The above algorithm finds an optimal $t$-round allocation.*

**Proof:** First, notice that the algorithm outputs a $t$-round allocation. Let us prove that it outputs an optimal one. Let $O = (o_1, ..., o_n)$ be an optimal $t$-round allocation, let $T$ be the set of bidders from Definition 3.1, and let $l = \Sigma_{i \in T} o_i$. For each bidder $i \in T$ remove the maximal number $q_i$ (possibly zero) of items from $o_i$ such that $v_i(o_i) = v_i(o_i - q_i)$. Observe that there exists a pair $(q'_j, p'_j)$ in $i$'s XOR bids such that $q'_j = o_i - q_i$. We now handle the bidders that are not in $T$. Each bidder $i \notin T$ holds a bundle that is a multiple of $b = \max(\left\lfloor \frac{m-l}{(n-t)^2} \right\rfloor, 1)$ in $O$, while in order for the allocation that we construct to be $t$-round we need the bidders not in $T$ to receive multiples of $b' = \max(\left\lfloor \frac{m-l'}{(n-t)^2} \right\rfloor, 1)$, for $l' = \Sigma_{i \in T}(o_i - q_i)$. However, notice that $b' \geq b$, and that the number of equi-size bundles is at least the same. Hence, by assigning each bidder $i \notin T$ the same number of equi-size bundles as in $O$, bidder $i$ holds at least the same value as in $O$. The lemma follows since the algorithm considers the newly constructed allocation. $\square$

We therefore have the following theorem:

**Theorem 3.5** *There exists a truthful VCG-based mechanism that provides a $(1 - \frac{1}{t+1})$-approximation for multi-unit auctions with $k$-minded bidders in time polynomial in $n$, $\log m$, $k$, for every constant $t$.*

## 3.2 A $\frac{1}{2}$-Approximation for Multi-Unit Auctions with Black-Box Access

Let us consider the multi-unit auction problem with general valuations given by black boxes. We will assume in our algorithm an "oracle access" to it that may be queried for $v_i(q)$, where $q$ is the given bundle size[4].

---

4. This is analogous to the weakest "value query" in a combinatorial auction setting. Our lower bounds presented later will apply to all other query types as well.

We will design a $\frac{1}{2}$-approximation MIR algorithm for this problem, which again yields an incentive-compatible VCG-based mechanism. Our MIR approximation algorithm will first split the items into $n^2$ equi-sized bundles of size $b = \lfloor \frac{m}{n^2} \rfloor$ as well as a single extra bundle of size $r$ that holds the remaining elements (thus $n^2 b + r = m$). The maximum in range algorithm will optimally allocate these whole bundles among the $n$ bidders. What we need to show are the following two simple facts:

**Lemma 3.6** *An optimal allocation of the bundles can be found in time polynomial in $n$ and $\log m$.*

**Lemma 3.7** *Let $(a_1, ..., a_n)$ be an optimal allocation of the bundles that was found by the algorithm, and $(o_1, ..., o_n)$ an optimal unrestricted allocation, then $\Sigma_i v_i(o_i) \leq 2\Sigma_i v_i(a_i)$.*

The proofs are simple:

**Proof:** (of Lemma 3.6): The algorithm is by dynamic programming. We calculate the following information for every $1 \leq i \leq n$ and $1 \leq q \leq n^2$: $M(i, q)$ is the maximum value that can be obtained by allocating at most $q$ regular bundles among bidders $1...i$, and $M^+(i, q)$ is the maximum value that can be obtained by allocating at most $q$ regular bundles and the "remainder" bundle among bidders $1...i$. Each entry can be filled in polynomial time using the relations: $M(i, q) = \max_{q' \leq q} v_i(q'b) + M(i - 1, q - q')$ and $M^+(i, q) = \max(\max_{q' \leq q}(v_i(q'b) + M^+(i - 1, q - q')), \max_{q' \leq q}(v_i(q'b + r) + M(i - 1, q - q')))$. $\qquad \square$

**Proof:** (of Lemma 3.7): Let us start with an optimal unrestricted allocation $o_1...o_n$ where all items are allocated (without loss of generality since the valuations are monotone), and look at the bidder $j$ that got the largest number of items $o_j \geq m/n$. There are now two possibilities: if $v_j(o_j) \geq \Sigma_{i \neq j} v_i(o_i)$ then by allocating all items to $j$ (i.e. all regular-sized bundles as well as the remainder bundle) we get the required 2-approximation. Otherwise, round *up* each $o_i$ to the nearest multiple of $b$ (i.e. to full bundles), except for bidder $j$ that gets nothing. This is a valid allocation since we added at most $nb \leq m/n$ items by rounding up, but deleted at least $m/n$ items by removing $o_j$, and its value is certainly at least $\Sigma_{i \neq j} v_i(o_i)$ which gives the required approximation. $\qquad \square$

We have thus proved:

**Theorem 3.8** *There exists a truthful polynomial time VCG-based mechanism that gives a $\frac{1}{2}$-approximation for multi-unit auctions with general valuations.*

## 4. Lower Bounds for VCG-Based Mechanisms

We now move on to show that both mechanisms essentially achieve the best approximation ratios possible. We say that an allocation $(s_1, ..., s_n)$ is *complete* if all items are allocated: $\Sigma_i s_i = m$. Consider an MIR algorithm for $n$ bidders that does not have full range of complete allocations. I.e., for some $0 \leq s_1, ..., s_{n-1} \leq m, \Sigma_{i<n} s_i \leq m$ it never outputs the allocation $(s_1, ..., s_{n-1}, m - \Sigma_{i<n} s_i)$. Now consider the set of valuations where for every bidder $i$ $v_i(q) = 1$ if and only if $q \geq s_i$ (and 0 otherwise). The only allocation with value $n$ is $(s_1, ..., s_{n-1}, m - \Sigma_{i<n} s_i)$ which is not in the range, while all other allocations have a value of at most $n - 1$.

From this we can easily get a lower bound for any computationally efficient MIR algorithm in the models considered in this paper. We start with a lower bound in the black-box model. The lower bound is on the number of queries that the bidders must be queried, and holds for any type of query – i.e., in a general communication setting.

**Proposition 4.1** *Let A be a MIR algorithm for multi-unit auctions that achieves an approximation ratio better than $\frac{1}{2}$. Then, the communication complexity of A is $\Omega(m)$.*

**Proof:** In the case of two bidders, an optimal algorithm is known to have a communication complexity of $\Theta(m)$:

**Lemma 4.2 (Nisan & Segal, 2006)** *Finding the optimal allocation in multi-unit auctions requires $\Omega(m)$ bits of communication, even if there are only two bidders and even for just finding the value of the allocation. This lower bound also applies to nondeterministic settings.*

Thus, any MIR algorithm for 2 bidders that uses $o(m)$ bits of communication will be non-optimal and thus, as argued above, gives no better than a $\frac{1}{2}$-approximation. The case of more than 2 bidders follows by setting all valuations but the first two to 0, and then considering all allocations in which all items are allocated to the first two bidders. □

The second result rules out the existence of FPTAS for $k$-minded bidders. In other words, the dependence of the running time in $\frac{1}{\epsilon}$ cannot be made polynomial. The result essentially applies to all models that allow single-minded bidders, e.g., XOR bids, and the bidding language used by Kothari et al. (2005) (see the next section for a description).

**Proposition 4.3** *Let A be a MIR algorithm that achieves an approximation ratio better than $(1 - \frac{1}{n})$. Then, A does not run in polynomial time, unless $P = NP$.*

**Proof:** Similarly to the previous proposition, and by the standard reduction from knapsack to multi-unit auctions, it follows that for every polynomial-time MIR algorithm there exist large enough $n$ and $m$ for which the range of complete allocations is not full, unless $P = NP$. The lemma follows by the discussion above. □

This concludes the proof of the lower bounds for MIR mechanisms. To draw the same conclusion for VCG-based mechanisms, one technical detail that should be explicitly mentioned. Our lower bounds were for MIR algorithms, while VCG-based mechanisms are only proved to give algorithms that are *equivalent* to MIR algorithms. See Dobzinski and Nisan (2007). However, both proofs hold even for finding the *value* of the optimal allocation and thus directly apply also to algorithms that are equivalent to MIR algorithms.

## 5. A General Construction and Applications

We present a general construction that takes a maximal-in-range algorithm for a constant number of bidders in some bidding language or model[5], and extends it to a truthful mechanism for an unbounded number of bidders. Yet, the extension loses only an arbitrarily small constant in the approximation ratio.

---

5. By a *model* we mean, e.g., some restriction on the valuations or on how they can be accessed.

We describe four applications of the construction. First, we reprove the PTAS for $k$-minded bidders and the $\frac{1}{2}$-approximation algorithm for general bidders of Section 3. Then, we study the bidding language considered by Kothari et al. (2005). Kothari et al. describe an *approximately* truthful FPTAS for this bidding language, while we present a truthful VCG-based PTAS (this is the best possible since Section 4 essentially rules out the possibility of a VCG-based FPTAS). Finally, we present a truthful $(\frac{3}{4}+\epsilon)$-approximation mechanism for the case the valuations are sub-additive (a.k.a. complement free) and are accessed via a black box.

## 5.1 The Setting

Fix some bidding language or a model for multi-unit auction in which the bidders can answer value queries. Let $A$ be a maximal-in-range algorithm for $t$ bidders and at most $m$ items in this model. Denote the complexity of $A$ by $A(t, m)$, its range by $\mathcal{R}_{A,t,m}$, and its approximation guarantee by $\alpha \leq 1$.

**The Construction**

1. Build the set $Q$ of allocations as follows:

   (a) Let $u = (1 + \frac{1}{2n})$.
       Let $L = \{0, 1, \lfloor u \rfloor, \lfloor u^2 \rfloor, \ldots, u^{\lfloor \log_u m \rfloor}, m\}$.

   (b) For every set $T$ of bidders, $|T| \leq t$, and $l \in L$:

       i. Run $A$ with $m - l$ items and the set $T$ of bidders. Denote by $s_i$ the number of items $A$ allocates to each bidder $i \in T$.
       ii. Split the remaining $l$ items into at most $2n^2$ bundles, each consisting of $\max(\lfloor \frac{l}{2n^2} \rfloor, 1)$ items.
       iii. Find the optimal allocation of the equal-size bundles among the bidders that are not in $T$. Denote by $s_i$ the allocation to each bidder $i \notin T$.
       iv. Add $(s_1, \ldots, s_n)$ to $Q$.

2. Output the allocation with the highest welfare in $Q$.

**Theorem 5.1** *There exists some range of allocations $\mathcal{R}$ such that the construction is maximal in $\mathcal{R}$. The construction runs in time $poly(\log m, n, A(t, m))$ for every constant $t$. It outputs an allocation with value of $(\alpha - \frac{1}{t+1})$ of the optimal allocation.*

**Proof:**   We will make use of the following definition:

**Definition 5.2** *An allocation is $(\mathcal{R}, t, l)$-round if:*

- *$\mathcal{R}$ is a set of allocations, and in each $R \in \mathcal{R}$ at most $t$ bidders are allocated non-empty bundles. The bidders are allocated together up to $m - l$ items.*

- *There exists a set $T$ of bidders, $|T| \leq t$, such that the bidders in $T$ are allocated according to some allocation in $\mathcal{R}$.*

- *Each bidder $i \notin T$ receives an exact multiple of $\max(\lfloor \frac{l}{2n^2} \rfloor, 1)$ units, and $\Sigma_{i \notin T} s_i \leq \max(\lfloor \frac{l}{2n^2} \rfloor, 1) \cdot n^2$.*

It is not hard to verify that the construction always outputs allocations that are in the following range:

$$\mathcal{R} = \{\mathcal{S} | S \text{ is a } (\mathcal{R}_{A,k,m-l}, k, l)\text{-round allocation where } l \in L \text{ and } k \leq t\}$$

**Lemma 5.3** *Let $(o_1, \ldots, o_n)$ be an optimal unrestricted allocation. There exists an allocation $(s_1, \ldots, s_n) \in \mathcal{R}$ for which $\Sigma_i v_i(s_i) \geq (\alpha - \frac{1}{t+1}) \cdot \Sigma_i v_i(o_i)$.*

**Proof:** Denote the value of the optimal solution by OPT. Without loss of generality, assume that all items are allocated and that $v_1(o_1) \geq \ldots \geq v_n(o_n)$. Let $l \in L$ be the largest such that $m - l \geq \Sigma_{i=1}^{t} o_i$. Let $(s_1, \ldots, s_t)$ be the allocation that $A$ outputs when run on bidders $1, \ldots, t$ and $m - l$ items, and assign each bidder $1 \leq i \leq t$ $s_i$ items. Observe that $\Sigma_{i=1}^{t} v_i(s_i) \geq \alpha \cdot \Sigma_{i=1}^{t} v_i(o_i)$. To finish the proof we show that there is an allocation in the range that recovers the value of all bidders $t+1, \ldots, n$ but at most one. The lemma will then follow since for each such bidder $i$, $v_i(o_i) \leq \frac{OPT}{t+1}$.

**Claim 5.4** *Step 1(b)iii returns an allocation $(s_{t+1}, \ldots, s_t)$ of the $l$ items to bidders $t + 1, \ldots, n$ such that for each bidder $i$ in this set, but at most one, we have that $v_i(s_i) \geq v_i(o_i)$.*

**Proof:** Let $r = \Sigma_{i=t+1}^{n} o_i$. Let $j \geq t + 1$ be some bidder with $o_j \geq \frac{r}{n}$ (observe that the existence of this bidder is guaranteed). Let $l \in L$ be the largest such that $m - l \geq \Sigma_{i=1}^{t} o_i$ (observe that $l \leq r$). We also have that $l \geq r - \frac{r}{2n}$: we chose the largest possible value for $l$, and therefore $l \geq \frac{r}{(1 + \frac{1}{2n})} \geq r - \frac{r}{2n}$.

Now, for each $i \neq j$ round up each $o_i$ to the nearest multiple of $\max(\lfloor \frac{l}{2n^2} \rfloor, 1)$, and allocate no items to bidder $j$. Observe that for all bidders but bidder $j$ the bundle size they get increases. Also observe that the number of additional items we allocate to bidders in $\{t+1, \ldots n\}$ is at most $\frac{l}{2n^2} \cdot n = \frac{l}{2n}$. Thus, we have that $l \geq r - \frac{r}{2n} + \frac{l}{2n} \geq r - o_j + \frac{l}{2n}$. $\square$

$\square$

All that is left is to show that the construction runs in polynomial time:

**Lemma 5.5** *The optimal allocation in $\mathcal{R}$ can be found in time $poly(\log m, n, A(t,m))$, for every constant $t$.*

**Proof:** Step 1(b)iii of the construction can be implemented using a dynamic programming similarly to Lemma 3.3; optimality of the allocation in $\mathcal{R}$ is clear. The running time is $poly(\log_{1 + \frac{1}{2n+1}} m \cdot n^k \cdot A(t,m))$, which is polynomial in the relevant parameters for every constant $t$. $\square$

$\square$

## 5.2 Applications of the Construction

We now provide several applications of our construction.

### 5.2.1 A PTAS for $k$-Minded Valuations

We reprove the PTAS for $k$-minded bidders of Section 3: a multi-unit auction problem with $m$ items and $t$ $k$-minded bidders can be optimally solved by exhaustive search in time $poly((tk)^t, \log m)$, which is polynomial in $\log m$ and $k$ for every constant $t$. By the construction (and since an optimal algorithm is in particular maximal in range), we get a PTAS for $k$-minded bidders: for every constant $t$, we get a $(1 - \frac{1}{t+1})$-approximation in time polynomial in $n$ and $\log m$.

### 5.2.2 A $\frac{1}{2}$-Approximation for General Valuations

Here we observe that a multi-unit auction with one bidder can be optimally solved by allocating all items to the single bidder. We let $t = 1$ in the statement of Theorem 5.1, and get a $\frac{1}{2}$-approximation algorithm for general valuations.

### 5.2.3 A PTAS for the Marginal Piecewise Bidding Language

The following marginal piecewise bidding language was used by Kothari et al. (2005): a valuation $v$ is determined by a list of at most $k$ tuples denoted by $(u_1, m_1), \ldots, (u_k, m_k)$. We assume that the $m_i$'s are non-negative and that $u_k > \ldots > u_1 = 1$. The tuples determine the marginal utility of the $j$th item. In other words, to determine the value of a set of $s$ items, we sum over all the marginal utilities. I.e., for each item $j$, $u_l \leq j < u_{l+1}$, let his marginal utility be $r_j = m_l$, and for every $s \leq m$, let $v(s) = \Sigma_{j=1}^{s} r_j$. (In fact, the above bidding language is more powerful than the one described by Kothari et al. (2005), which allows only marginal-decreasing piecewise valuations.)

We now show how to optimally solve a multi-unit auction problem in this setting with a constant number of bidders. A PTAS follows, just as in the $k$-minded case.

We say that bidder $i$ is *precisely assigned* if he is allocated $s_i$ items, and for some $u_i > 0$ there exists a tuple $(s_i, u_i)$ in his $k$ bids. The main observation here is that there is an optimal solution $(o_1, \ldots, o_n)$ in which at most one bidder is not precisely assigned: suppose there are two bidders $i$ and $i'$ that are not precisely assigned. Then, move items to the bidder with the higher (or equal) marginal utility. The value of the allocation cannot decrease. Continue this process until all bidders but at most one are precisely assigned.

Now optimally solving a multi-unit auction problem with a constant number of bidders is obvious: select each of the $n$ bidders in his turn to be the bidder that is not precisely assigned. In each iteration, let $i$ be the bidder that is not precisely assigned, and go over all allocations in which all other bidders are precisely assigned. Then, assign bidder $i$ the remaining items. Since there are at most $k$ possible sets that make a bidder precisely assigned, the algorithm runs in time $poly(n \cdot (n \cdot k)^{n-1}, \log m)$, which is polynomial in $\log m$ and $k$ for every constant $n$.

### 5.2.4 A $(\frac{3}{4} + \epsilon)$-Approximation for Subadditive Valuations

In this model we assume that the valuations are given as black boxes (as in Section 3.2), and that for each valuation $v$ and bundles $0 \leq s, t \leq m$ we have that $v(s) + v(t) \geq v(s+t)$. Such valuations are called *subadditive* valuations.

Let us now describe an algorithm that provides an approximation ratio of $\frac{3}{4}$ in this setting when the number of bidders is constant. By the construction, we get a $(\frac{3}{4} + \epsilon)$-approximation VCG-based mechanism for unbounded number of bidders, for every constant $\epsilon$. The algorithm is quite simple: Fix a small enough constant $\delta > 1$ ($\delta = \frac{4}{3}$ suffices). All bidders, but at most one, can only receive a bundle $s$ that is a power of $\delta$ (including the empty bundle). The bidder that does not get a bundle of size that is a power of $\delta$ receives the remaining items. We use exhaustive search to find the optimal allocation in this range.

To see that the algorithm indeed provides an approximation ratio of $\frac{3}{4}$, consider an optimal solution $(o_1, \ldots, o_n)$. Without loss of generality, assume that $o_1 \geq \ldots \geq o_n$ (notice that unlike before now the bidders are ordered by their bundle size). Let $O$ be the set of bidders with odd indices, and $E$ be the set of bidders with even indices.

The analysis is divided into two cases. First suppose that $\Sigma_{i \in O} v_i(o_i) \geq \Sigma_{i \in E} v_i(o_i)$. Consider the following allocation: the bundles of bidders in $E$ are rounded up to the power of $\delta$ *that is near* $\frac{o_i}{2}$, the bundles of bidders in $O \setminus \{1\}$ are rounded up to the nearest power of $\delta$, and bidder 1 gets the remaining items. Notice that the above allocation is valid since for a small enough choice of $\delta$ we assign bidders in $O$ no more items than what we removed from bidders in $E$. Also notice that this allocation is in the range. As for the approximation ratio, observe that bidders in $O$ hold at least the same value as in the optimal solution, since each bidder in $O$ is allocated at least the same number of items as in the optimal solution. In addition, each bidder in $E$ holds at least half of the items allocated to him in the optimal solution. Thus, by subadditivity, bidders in $E$ hold together at least half of the value they hold in the optimal solution. In total, the value of the allocation obtained by the algorithm is at least $\frac{3}{4}$ of the value of the optimal solution.

Let us now handle the case where $\Sigma_{i \in O} v_i(o_i) < \Sigma_{i \in E} v_i(o_i)$. Consider the allocation where all bidders in $O$ are rounded up to the power of $\delta$ that is near $\frac{o_i}{2}$, and all bidders in $E$ are rounded up to the nearest power of $\delta$ (except for one arbitrary bidder in $E$ who gets the remaining items). This allocation is in the range, and the analysis is similar to the previous case, leaving us with an approximation ratio of $\frac{3}{4}$ also in the current case.

The running time of the algorithm is $poly(n \cdot (\log_\delta m)^{n-1})$, which is polynomial in $\log m$ for constant $n$ and $\delta$.

Notice that the approximation ratio achieved is almost the best possible, as every MIR approximation algorithm that guarantees a factor better than $\frac{3}{4}$ requires $\Omega(m)$ communication: by Lemma 4.2 finding the optimum solution of a multi-unit auction with two bidders requires $\Omega(m)$ bits of communication. We make the valuations sub-additive by defining for each $v$ a new valuation: $v'(s) = v(s) + v(m)$, for all $s \neq 0$. Thus, as in Section 4, the range of every polynomial-time MIR mechanism for two bidders with subadditive valuations cannot yield all complete allocations. Fix some MIR algorithm, and let $(s_1, m - s_1)$ be a complete allocation that is not in the range. Consider the following instance: bidder 1 values at least $s_1$ items with a value of 2, and smaller bundles with a value of 1, and bidder 2 values at least $m - s_2$ items with a value of 2, and smaller bundles with a value of 1 (and 0 is the value of the empty bundle). Notice that the valuations of the bidders are indeed subadditive. Also observe that the optimal welfare is 4, but the mechanism can achieve welfare of at most 3.

## Acknowledgments

A preliminary version of this paper appeared in EC'07. We are grateful to Liad Blumrosen and Ahuva Mu'alem for helpful discussions. The second author is supported by a grant from the Israeli Science Foundation.